\documentclass[aps,pre]{revtex4}
\usepackage[T1]{fontenc}
\usepackage{graphicx}
\usepackage{amssymb,amsfonts,amsmath}
\usepackage{dsfont}
\usepackage{hyperref}
\usepackage{url}
\usepackage{dsfont}

\DeclareMathOperator{\atanh}{atanh}
\DeclareMathAlphabet{\varmathbb}{U}{bbold}{m}{n}

\newcommand{\eq}[1]{\begin{align}#1\end{align}}

\begin{document}

\title{Spectral Clustering of Graphs with the Bethe Hessian}

\author{A. Saade$^1$, F. Krzakala$^{1,2}$ and L. Zdeborov\'a$^3$}

\affiliation{ $^1$ Laboratoire de Physique Statistique, CNRS UMR 8550,
  Universit\'e P. et M. Curie Paris 6 et \'Ecole Normale Sup\'erieure,
  24, rue Lhomond, 75005 Paris, France.\\ $^2$ ESPCI and CNRS UMR 7083
  Gulliver, 10 rue Vauquelin,Paris 75005\\ $^3$ Institut de Physique
  Th\'eorique, CEA Saclay and URA 2306, CNRS, 91191 Gif-sur-Yvette,
  France}

\begin{abstract}
  Spectral clustering is a standard approach to label nodes on a graph
  by studying the (largest or lowest) eigenvalues of a symmetric real
  matrix such as e.g. the adjacency or the Laplacian. Recently, it has
  been argued that using instead a more complicated, non-symmetric and
  higher dimensional operator, related to the non-backtracking walk on
  the graph, leads to improved performance in detecting clusters,
  and even to optimal performance for the stochastic block model. Here,
  we propose to use instead a simpler object, a symmetric real matrix
  known as the Bethe Hessian operator, or deformed Laplacian. We show
  that this approach combines the performances of the non-backtracking
  operator, thus detecting clusters all the way down to the
  theoretical limit in the stochastic block model, with the
  computational, theoretical and memory advantages of real symmetric
  matrices.
\end{abstract}

\date{\today}
\maketitle

Clustering a graph into groups or functional modules (sometimes called
communities) is a central task in many fields ranging from machine
learning to biology.  A common benchmark for this problem is to
consider graphs generated by the stochastic block model (SBM)
\cite{Holland1983109,wang1987stochastic}. In this case, one considers
$n$ vertices and each of them has a group label $g_v \in
\{1,\ldots,q\}$. A graph is then created as follows: all edges are
generated independently according to a $q \times q$ matrix $p$ of
probabilities, with $\Pr[A_{u,v} = 1] = p_{g_u,g_v}$. The group labels
are hidden, and the task is to infer them from the knowledge of the
graph. The stochastic block model generates graphs that are a
generalization of the Erd\H{o}s-R\'enyi ensemble where an unknown
labeling has been hidden.

We concentrate on the sparse case, where algorithmic
challenges appear. In this case $p_{ab}$ is $O(1/n)$, and we
denote $p_{ab} = c_{ab}/n$. For simplicity we concentrate on the
most commonly-studied case where groups are equally sized, $c_{ab} = c_{\rm in}$ if $a = b$ and
$c_{ab}=c_{\rm out}$ if $a \ne b$.  Fixing $c_{\rm in}>c_{\rm out}$ is referred to
as the assortative case, because vertices from the same group connect
with higher probability than with vertices from other groups. $c_{out
}>c_{\rm in}$ is called the disassortative case. An important conjecture
\cite{decelle2011asymptotic} is that any tractable algorithm will
only detect communities if
\eq{
\lvert
c_{\rm in}-c_{\rm out}\rvert>q\sqrt{c}\, ,
\label{limit}
}
where $c$ is the average degree. In the case of $q\!=\!2$ groups, in
particular, this has been rigorously proven
\cite{mossel2013proof,massoulie2013community} (in this case, one
can also prove that no algorithm could detect communities if this
condition is not met). An ideal clustering algorithm should have a
low computational complexity while being able to perform optimally for
the stochastic block model, detecting clusters down to the transition (\ref{limit}).

So far there are two algorithms in the literature that are able to
detect clusters down to the transition (\ref{limit}). One is a message-passing algorithm
based on belief-propagation
~\cite{decelle2011inference,decelle2011asymptotic}. This algorithm, however, needs to be
fed with the correct parameters of the stochastic block model to perform well, and its computational complexity scales
quadratically with the number of clusters, which is an important
practical limitation. To avoid such problems, the most popular
non-parametric approach to clustering are spectral methods, where one
classifies vertices according to the eigenvectors of a matrix
associated with the network, for instance its adjacency matrix
\cite{Tutorial,newman2006finding}. However, while this works
remarkably well on regular, or dense enough graphs
\cite{bickel2009nonparametric}, the standard versions of spectral
clustering are suboptimal on graphs generated by the SBM, and in some cases completely fail
to detect communities even when other (more complex) algorithms such
as belief propagation can do so. Recently, a new class of spectral
algorithms based on the use of a non-backtracking walk on the directed
edges of the graph has been introduced in \cite{krzakala2013spectral}
and argued to be better suited for spectral clustering. In particular,
it has been shown that this operator is
optimal for graphs generated by the stochastic block model, and able
to detect communities even in the sparse case all the way
down to the theoretical limit (\ref{limit}).

These results are, however, not entirely
satisfactory. First, the use a of a high-dimensional matrix (of
dimension $2m$ - where $m$ is the number of edges - rather than $n$,
the number of nodes) can be
expensive, both in terms of computational time and memory. Secondly,
linear algebra methods are faster and more efficient for symmetric
matrices than non-symmetric ones. The first problem was partially
resolved in \cite{krzakala2013spectral} where an equivalent operator
of dimensions $2n$ was shown to exist. It was still, however, a non
symmetric one and more importantly, the reduction does not extend to
weighted graphs, and thus presents a strong limitation.

In this contribution, we provide the best of both worlds: a
non-parametric spectral algorithm for clustering with a symmetric,
real matrix that performs as well, and in fact slightly better, than
the non-backtracking operator of \cite{krzakala2013spectral}. This operator is actually not new, and has been known as
the Bethe Hessian in the context of statistical physics and machine
learning \cite{mooij2004validity,ricci2012bethe} or the deformed
Laplacian in various other fields. We show that the spectrum of
this operator is directly linked to that of the non-backtracking
matrix, that it also performs optimally for the stochastic block
model, in the sense that it identifies communities as soon as
(\ref{limit}) holds. It also performs well in clustering standard real
world benchmark networks.

The paper is organized as follows. In Sec.~\ref{sec:BH} we give the
expression of the Bethe Hessian operator. We discuss in detail its
properties and its connection with both 
the non-backtracking operator and an Ising spin glass in Sec.~\ref{sec:PW}.  In
Sec.~\ref{sec:CAVITY}, we study analytically the spectrum in the case
of the stochastic block model.   Finally, in
Sec.~\ref{sec:numerics} we perform numerical tests on both the
stochastic block model and on some real networks. 

\section{Clustering based on the Bethe Hessian matrix}
\label{sec:BH}

Let $\mathcal{G}=(V,E)$ be a graph with $n$ vertices,
$V=\{1,...,n\}$. Denote by $A$ its adjacency matrix, and by $D$ the
diagonal matrix defined by $D_{ii}=d_i,\ \forall i\in V$, where $d_i$
is the degree of vertex $i$.  We then define the Bethe Hessian matrix, 
sometimes called the deformed Laplacian, as \eq{
\label{BH}
H(r):=(r^2-1)\mathds{1}-rA+D\, , } where $\lvert r\rvert>1$ is a
regularizer that we will set to a well-defined value $\lvert r\rvert=r_c$ depending
on the graph, for instance $r_c=\sqrt{c}$ in the case of the
stochastic block model, where $c$ is the average degree of the
graph (see Sec.~\ref{sec:B}).  

The spectral algorithm that is the main result of this paper works as follows: we compute the eigenvectors associated with the negative eigenvalues
of both $H(r_c)$ and $H(-r_c)$, and cluster them with a standard
clustering algorithm such as k-means (or simply by looking at the sign
of the components in the case of two communities). The negative
eigenvalues of $H(r_c)$ reveal the assortative aspects, while
those of $H(-r_c)$ reveal the disassortative ones.

\begin{figure}[t]
\begin{center}
\includegraphics[width=0.99\linewidth]{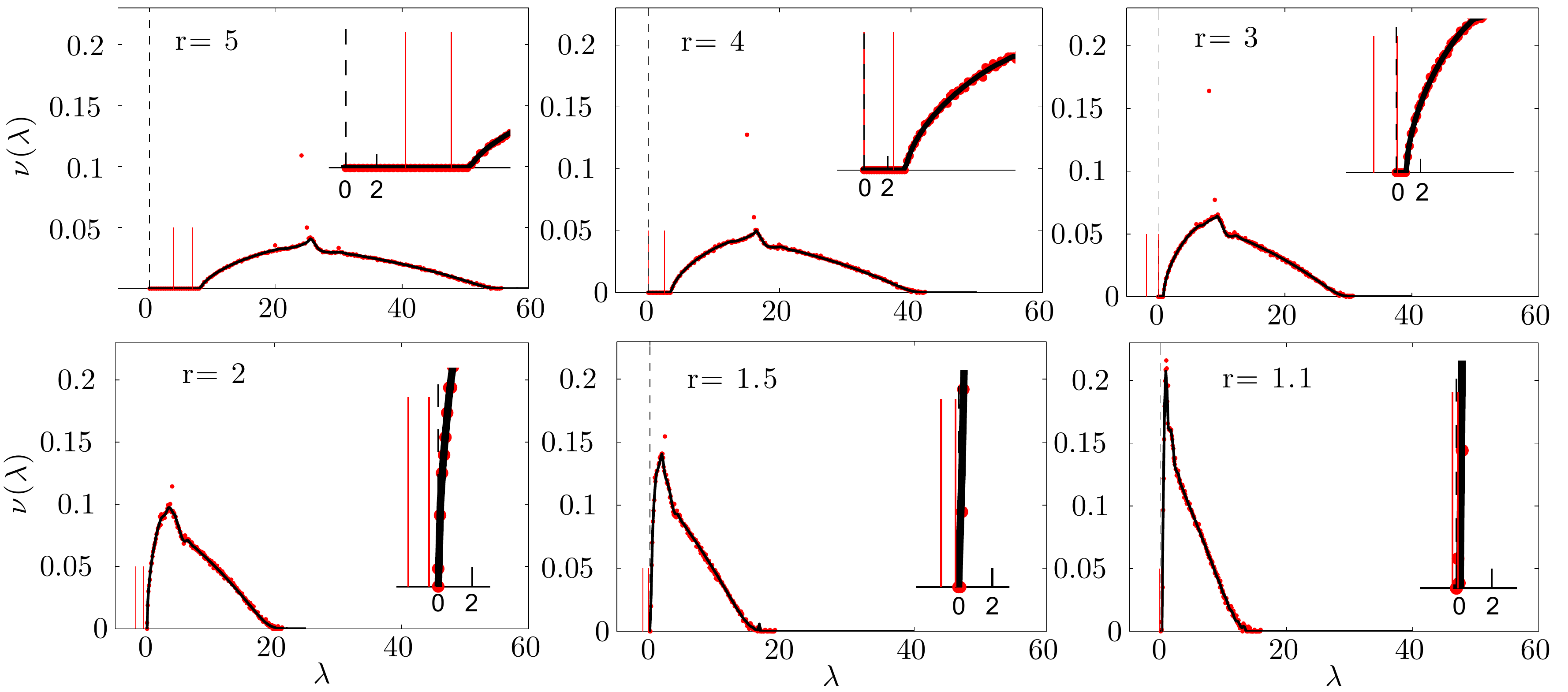} 	
\end{center}
\caption{Spectral density of the Bethe Hessian for various values of
  the regularizer $r$ on the stochastic block model.  The red dots are the result of the direct
  diagonalization of the Bethe Hessian for a graph of $10^4$ vertices
  with $2$ clusters, with $c\!=\!4, c_{\rm in}\!=\!7,
  c_{\rm out}\!=\!1$. 
The black curves
  are the exact solutions to the recursion (\ref{BPrec}) for
  $c\!=\!4$, obtained from population dynamics (with a population of
  size $10^5$), see section \ref{sec:CAVITY}.
We isolated the two smallest eigenvalues,
  represented as small bars for convenience. The dashed black line
  marks the $x\!=\!0$ axis, and the inset is a zoom around this axis.
  At large value of $r$ (top left) $r\!=\!5$, the Bethe Hessian is
  positive definite and all eigenvalues are positive. As $r$ decays,
  the spectrum moves towards the $x\!=\!0$ axis. The smallest
  (non-informative) eigenvalue reaches zero for $r\!=\!c\!=\!4$
  (middle top), followed, as $r$ decays further, by the second
  (informative) eigenvalue at
  $r\!=\!(c_{\text{\rm in}}-c_{\text{\rm out}})/2\!=\!3$, which is the value
  of the second largest eigenvalue of $\rm B$ in this case
  \cite{krzakala2013spectral} (top right). Finally, the bulk reaches
  $0$ at $r_c\!=\!\sqrt{c}\!=\!2$ (bottom left). At this point, the
  information is in the negative part, while the bulk is in the
  positive part. Interestingly, if $r$ decays further (bottom middle
  and right) the bulk of the spectrum remains positive, but the
  informative eigenvalues blend back into the bulk. The best choice is
  thus to work at $r_c\!=\!\sqrt{c}\!=\!2$. \label{fig:CAVITY}}
\end{figure}

Figure \ref{fig:CAVITY} illustrates the spectral properties of the
Bethe Hessian (\ref{BH}) for networks generated by the stochastic
block model.
When $r= \pm \sqrt{c}$ the informative eigenvalues (i.e. those having
eigenvectors correlated to the cluster structure) are
  the negative ones, while the non-informative bulk
  remains positive. 
There are as many
 negative eigenvalues as there are hidden clusters. It is thus
  straightforward to select the relevant eigenvectors.  This is very
  unlike the situation for the operators used in standard spectral
  clustering algorithms (except, again, for the non-backtracking
  operator) where one must decide in a somehow ambiguous way which
  eigenvalues are relevant (outside the bulk) or not (inside the
  bulk). Here, on the contrary, no prior knowledge of the number of
  communities is needed to select them.




On more general graphs, we argue that the best choice for the
  regularizer is $r_c=\sqrt{\rho({\rm B})}$, where $\rho({\rm B})$ is the spectral
  radius of the non-backtracking operator. We support this claim both
  numerically, on real world networks (sec.~\ref{sec:REAL}), and
  analytically (sec.~\ref{sec:CAVITY}). We also show that $\rho({\rm B})$ can be
  computed without building the matrix $\rm B$ itself, by efficiently
  solving a quadratic eigenproblem (sec.~\ref{sec:B}).

The Bethe Hessian can be generalized straightforwardly to the
  weighed case: if the edge $(i,j)$ carries a weight $w_{ij}$, then
  we can use the matrix $\tilde{H}(r)$ defined by 
 \eq{
\label{weighted}
\tilde{H}(r)_{ij}=\delta_{ij}\Big(1 + \underset{k\in\partial
  i}{\sum}\frac{w_{ik}^2}{r^2-w_{ik}^2}\Big)-\frac{rw_{ij}A_{ij}}{r^2-w_{ij}^2}\,
,}
where $\partial i$ denotes the set of neighbors of vertex $i$. This is in fact the general expression of the Bethe Hessian of a certain weighted statistical model (see section \ref{section:MODEL}).
 If all weights are equal to unity, $\tilde{H}$ reduces to (\ref{BH}) up to a trivial factor. Most of the arguments developed in the following generalize immediately to $\tilde{H}$, including the relationship with the \emph{weighted} non-backtracking operator, introduced in the conclusion of \cite{krzakala2013spectral}.

%

\section{Derivation and relation to previous works}
\label{sec:PW}
Our approach is connected to both the
spectral algorithm using the non-backtracking matrix and to an Ising
spin glass model. We now
discuss these connections, and the properties of the Bethe
Hessian operator along the way.

\subsection{Relation with the non-backtracking matrix}
\label{sec:B}
In this section we describe the relationship between the spectrum of the Bethe Hessian
and that of the non-backtracking operator of
\cite{krzakala2013spectral} defined as a $2M \times 2M$ non-symmetric
matrix indexed by the directed edges of the graph $i\to j$ 
  \eq{
\label{eq:B} 
    {\rm B}_{i\to j,k\to l} = \delta_{jk} (1-\delta_{il}) \, . 
}
The remarkable efficiency of the non-backtracking operator
is due to the particular structure of its (complex) spectrum. For
graphs generated by the SBM the spectrum
decomposes into a bulk of uninformative eigenvalues sharply
constrained when $n\!\rightarrow\!\infty$ to the disk of radius
$\sqrt{\rho({\rm B})}$, where $\rho({\rm B})$ is the spectral radius of $\rm B$
\cite{saade2014spectral}, well separated from the real, informative
eigenvalues, that lie outside of this circle.  It was also remarked
that the number of real eigenvalues outside of the circle is the number of
communities, when the graph was generated by the stochastic
block model. More precisely, the presence of assortative communities
yields real positive eigenvalues larger than $\sqrt{\rho({\rm B})}$, while
the presence of disassortative communities yields real negative
eigenvalues smaller than $-\sqrt{\rho({\rm B})}$.  The authors of
\cite{krzakala2013spectral} showed that all eigenvalues $\lambda$ of
$\rm B$ that are different from $\pm1$ are roots of the polynomial \eq{
\label{Ihara-Bass}
\det{[(\lambda^2-1)\mathds{1}-\lambda A+D]}=\det{H(\lambda)}\, . } This is
known in graph theory as the Ihara-Bass formula for the graph zeta
function. It provides the link between $\rm B$ and the (determinant of
the) Bethe Hessian (already noticed in
\cite{watanabe2009graph}): a real eigenvalue of $\rm B$ corresponds to a
value of $r$ such that the Bethe Hessian has a vanishing eigenvalue.


For any finite $n$, when $r$ is large enough, $H(r)$ is positive
definite. This can be proven by using e.g. the Gershgorin circle
theorem. Then as $r$ decreases, a new negative eigenvalue of
$H(r)$ appears when it crosses the zero axis, i.e whenever $r$ is
equal to a real positive eigenvalue of $\rm B$. Of course, the same
phenomenon takes place when increasing $r$ from a large negative
value. In order to translate all the informative eigenvalues of $\rm
B$ into negative eigenvalues of $H(r)$ we will adopt \eq{
\label{fixingReg}
r_c=\sqrt{\rho({\rm B})} \, .} 
since all the relevant values of $\rm B$ are {\it outside} the circle
of radius $r_c$. Note also that since the eigenvalues of $\rm B$ come in
pairs having their product close to $\rho({\rm B})$ \cite{krzakala2013spectral}, for $r<r_c$ the
negative eigenvalues of $H(r_c)$ move back into the positive part of
the spectrum. Hence taking $r<r_c$ is not desirable.

Let us stress that to compute $\rho({\rm
  B})$, we do not need to actually build the non-backtracking
matrix. First, when the degree to degree correlations are small
\cite{krzakala2013spectral}, $\rho({\rm B})=\langle
d^2\rangle/\langle d\rangle-1$. In a more general setting, we can
efficiently refine this initial guess by solving for the closest root
of the quadratic eigenproblem defined by (\ref{Ihara-Bass}),
e.g. using a standard SLP algorithm \cite{ruhe1973algorithms}. This
approach is implemented for the real world networks of
sec.~\ref{sec:REAL}. With the choice (\ref{fixingReg}), the
informative eigenvalues of $\rm B$ are in one-to-one correspondance
with the union of negative eigenvalues of $H(r_c)$ and $H(-r_c)$. Their
number will therefore tell us the number of (detectable) communities
in the graph, and we will use them to infer the community membership
of the nodes, by using a standard clustering algorithm such as
k-means. In particular, the Bethe Hessian detects communities whenever
$\rm B$ does, that is down to the theoretical threshold
\cite{krzakala2013spectral}.

\subsection{Hessian of the Bethe free energy}
\label{section:MODEL}
Let us define a pairwise Ising model on the graph $\mathcal{G}$ by the joint probability
distribution:
 \eq{
\label{graphModel}
P(\{x\})=\frac{1}{Z}\exp{\left( \underset{(i,j)\in E}\sum
    \atanh\Big(\frac{1}{r}\Big)x_ix_j\right) }\, , } 
where $\{x\}:=\{x_i\}_{i\in \{1..n\}}\in\{\pm1\}^n$is a set of binary
random variables sitting on the nodes of the graph $\mathcal{G}$. The
regularizer $r$ is here a parameter that controls the strength of
the interaction between the variables: the larger $\lvert r\rvert$
is, the weaker is the interaction (this would be analogous to the
temperature in a statistical physics).

In order to study this model, a standard approach in machine learning
is the Bethe approximation \cite{WJ2008} in which the means $\langle
x_i\rangle$ and moments $\langle x_ix_j\rangle$ are approximated by
the parameters $m_i$ and $\xi_{ij}$ that minimize the so-called Bethe
free energy $F_{\rm Bethe}(\{m_i\},\{\xi_{ij}\})$ defined as
\eq{
\label{freeEnergy}
\notag F_{\rm Bethe}(\{m_i\},\{\xi_{ij}\})&=-\underset{(i,j)\in E}{\sum}\atanh\Big(\frac{1}{r}\Big)\xi_{ij}+\underset{(i,j)\in E}{\sum}\ \underset{x_i,x_j}{\sum}\eta\Big(\frac{1+m_ix_i+m_jx_j+\xi_{ij}x_ix_j}{4}\Big)\\
&+\underset{i\in
  V}{\sum}(1-d_i)\underset{x_i}{\sum}\eta\Big(\frac{1+m_ix_i}{2}\Big)\, ,
}
where $\eta(x):=x\ln{x}$.
Such approach allows for instance to derive the belief propagation ($\rm BP$)
algorithm. Here, however, we
wish to restrict to a spectral algorithm.

At very high $r$ the minimum of the Bethe free energy is given by
the so-called  paramagnetic point $m_i=0,\ \xi_{ij}=\frac{1}{r}$. It turns out
\cite{mooij2004validity} that the $m_i=0,\ \xi_{ij}=\frac{1}{r}$ is a stationarity
point of the Bethe free energy for every $r$. 
Instead of considering the complete Bethe free energy, we
will consider only its behavior around the paramagnetic point. This
can be expressed via the Hessian (matrix of second derivatives), that
has been studied extensively, see
e.g.~\cite{mooij2004validity}, \cite{ricci2012bethe}. 
At the paramagnetic point, the blocks of
the Hessian involving one derivative with respect to the $\xi_{ij}$ are $0$, and the block involving two such derivatives is a positive definite diagonal matrix \cite{watanabe2009graph}. We will therefore, somewhat
improperly, call Hessian the matrix \eq{
  \mathcal{H}_{ij}(r)=\frac{\partial F_{\rm Bethe}}{\partial
    m_i\partial m_j}\Big\lvert_{m_i=0,\xi_{ij}=\frac{1}{r}}\, . } In
particular, at the paramagnetic point: \eq{
  \mathcal{H}(r)=\mathds{1}+\frac{D}{r^2-1} -
  \frac{rA}{r^2-1}=\frac{H(r)}{r^2-1} \, .}
A more general expression of the Bethe Hessian in the case of weighted
interactions $\atanh(w_{ij}/r)$ (with weights rescaled to be in
$[0,1]$) is given by eq. (\ref{weighted}).  All eigenvectors of $H(r)$ and
$\mathcal{H}(r)$ are the same, as are the eigenvalues up to a multiplicative,
positive factor (since we consider only $|r|>1$). 

The paramagnetic point is stable if and only if
$H(r)$ is positive definite. The appearance of each negative
eigenvalue of the Hessian corresponds to a phase transition in the Ising model at which a new cluster
(or a set of clusters) starts to be identifiable. The corresponding
eigenvector will give the direction towards the cluster labeling. This motivates the use of the Bethe
Hessian for spectral clustering. 

For tree-like graphs such as those generated by the
SBM, model (\ref{graphModel}) can been studied
analytically in the asymptotic limit $n\!\to\! \infty$. 
The location of the possible phase transitions in
model (\ref{graphModel}) are also known from spin glass theory and the
theory of phase transitions on random graphs (see
e.g.~\cite{mooij2004validity,decelle2011inference,decelle2011asymptotic,ricci2012bethe}). For
positive $r$ the trivial ferromagnetic phase appears at $r=c$, while
the transitions towards the phase corresponding to the hidden community
structure arise between $\sqrt{c}\! <\! r \!<\! c$.  For disassortative communities,
the situation is symmetric with $r \!<\!-\sqrt{c}$.  Interestingly, at $r=\pm
\sqrt{c}$, the model undergoes a spin glass phase transition. At this point all the relevant eigenvalues have passed in
the negative side (all the possible transitions from the paramagnetic
states to the hidden structure have taken place) while the bulk on
non-informative ones remains positive. This scenario is illustrated in
Fig.~\ref{fig:CAVITY} for the case of two assortative clusters.

\section{The spectrum of the Bethe Hessian}
\label{sec:CAVITY}
In this section, we show how the spectral density of the Bethe
Hessian can be computed analytically on tree-like graphs such as those
generated by the stochastic block model. This will serve two goals: i)
to justify independently our choice for the value of the regularizer $r$
and ii) to show that for all values of $r$, the bulk of uninformative
eigenvalues remains in the positive region.  The spectral density is
defined by: \eq{
  \nu(\lambda)=\frac{1}{n}\overset{n}{\underset{i=1}{\sum}}\
  \delta(\lambda-\lambda_i)\, , } where the $\lambda_i$'s are the
eigenvalues of the Bethe Hessian. It can be shown
\cite{rogers2008cavitySym} that the spectral density (in
which potential delta peaks have been removed) is given by \eq{
\label{specDens}
\nu(\lambda)=\frac{1}{\pi n}\overset{n}{\underset{i=1}{\sum}}\
\text{Im}\Delta_i(\lambda)\, ,
}
where the $\Delta_i$ are complex variables living on the vertices of the graph $\mathcal{G}$, which are given by:
\eq{
\label{full}
\Delta_i=\Big(-\lambda+r^2+d_i-1-r^2\underset{l\in \partial
  i}{\sum}\Delta_{l\rightarrow i}\Big)^{-1}\, ,
}
where $d_i$ is the degree of node $i$ in the graph, and $\delta_i$ is
the set of neighbors of $i$. The $\Delta_{i\rightarrow j}$ are the
(linearly stable) solution of the following belief propagation
recursion, or cavity method \cite{MM2009}, 
\eq{
\label{BPrec}
\Delta_{i\rightarrow
  j}=\Big(-\lambda+r^2+d_i-1-r^2\underset{l\in \partial i\backslash
  j}{\sum}\Delta_{l\rightarrow i}\Big)^{-1}\, . } The ingredients to
derive this formula are to turn the computation of the spectral
density into a marginalization problem for a graphical model on the
graph $\mathcal{G}$, and then write the belief propagation equations to solve
it. It can be shown \cite{RSA:RSA20313} that this approach leads to an
asymptotically exact description of the spectral density on random graphs such
as those generated by the stochastic block model, which are locally
tree-like in the limit where $n\rightarrow\infty$. We can solve
equation (\ref{BPrec}) numerically using a population dynamics algorithm
\cite{MM2009}: starting from a pool of variables, we iterate by
drawing at each step a variable, its excess degree and its neighbors
from the pool, and updating its value according to (\ref{BPrec}). The
results are shown on Fig.~\ref{fig:CAVITY}: the bulk of the spectrum is
always positive.

We now justify analytically that the bulk of eigenvalues of the Bethe
Hessian reaches $0$ at $r=\sqrt{\rho({\rm B})}$. From (\ref{specDens}) and
(\ref{full}), it is clear that if the linearly stable solution of
(\ref{BPrec}) is real, then the corresponding spectral density will be
equal to $0$. We want to show that there exists an open set
$U\subset\mathbb{R}$ around $0$ in which there exists a real, stable,
solution to the BP recursion.  Let us call $\underline{\Delta}\in
\mathbb{R}^{2M}$, where $M$ is the number of edges in $\mathcal{G}$,
the vector which components are the $\Delta_{i\to j}$. We introduce
the function $F:
(\lambda,\underline{\Delta})\in\mathbb{R}^{2M+1}\rightarrow
F(\lambda,\underline{\Delta})\in\mathbb{R}^{2M}$ defined by \eq{
\label{F}
F(\lambda,\underline{\Delta})_{i\to
  j}=\Big(-\lambda+r^2+d_i-1-r^2\underset{l\in \partial i\backslash
  j}{\sum}\Delta_{l\rightarrow i}\Big)-\frac{1}{\Delta_{i\rightarrow
    j}}\, ,
}
so that equation (\ref{BPrec}) can be rewritten as
\eq{
\label{fixedPoint}
F(\lambda,\underline{\Delta})=0\, . } It is straightforward to check that when
$\lambda=0$, the assignment $\Delta_{i\rightarrow j}=1/r^2$ is a real
solution of (\ref{fixedPoint}). Furthermore, the Jacobian of $F$ at
this point reads  \eq{ J_{F}(0,\{1/r^2\})=\begin{pmatrix}
    -1 &    \\
    0 &  \\
     \vdots  & &$$ \mbox{\Large $ {r^2(r^2\mathds{1}-B)}$ } $$ \\
   & \\
    0 &
 \end{pmatrix}\, ,
} where $\rm B$ is the $2M\times2M$ non-backtracking operator and
$\mathds{1}$ is the $2M\times2M$ identity matrix. The square submatrix of the Jacobian containing the derivatives with respect to the messages $\Delta_{i\to j}$ is
therefore invertible whenever $r>\sqrt{\rho({\rm B})}$. From the
continuous differentiability of $F$ around $(0,\{1/r^2\})$ and the implicit function
theorem, there exists an open set $V$ containing $0$ such that for all
$\lambda\in V$, there exists
$\tilde{\underline{\Delta}}(\lambda)\in\mathbb{R}$ solution of
(\ref{fixedPoint}) , and the function $\tilde{\underline{\Delta}}$ is
continuous in $\lambda$. To show that the spectral density is indeed 0
in an open set around $\lambda=0$, we need to show that this solution
is linearly stable. Introducing the function
$G_{\lambda}:\underline{\Delta}\in \mathbb{R}^{2M}\rightarrow
G_{\lambda}(\underline{\Delta})\in\mathbb{R}^{2M}$ defined by \eq{
  G_{\lambda}(\underline{\Delta})_{i\rightarrow
    j}=\Big(-\lambda+r^2+d_i-1-r^2\underset{l\in \partial i\backslash
    j}{\sum}\Delta_{l\rightarrow i}\Big)^{-1} } it is enough to show
that the Jacobian of $G_{\lambda}$ at the point
$\tilde{\underline{\Delta}}(\lambda)$ has all its eigenvalues smaller
than 1 in modulus, for $\lambda$ close to $0$. But since
$J_{G_{\lambda}}(\underline{\Delta})$ is continuous in
$(\lambda,\underline{\Delta})$ in the neighborhood of
$(0,\tilde{\underline{\Delta}}(0)=\{1/r^2\})$, and
$\tilde{\underline{\Delta}}(\lambda)$ is continuous in $\lambda$, it
is enough to show that the spectral radius of $J_{G_0}(\{1/r^2\})$ is
smaller than $1$. We compute \eq{ J_{G_0}(\{1/r^2\})=\frac{1}{r^2}{\rm
    B} } so that the spectral radius of $J_{G_0}(\{1/r^2\})$ is
$\rho({\rm B})/r^2$, which is (strictly) smaller than $1$ as long as
$r>\sqrt{\rho({\rm B})}$. From the continuity of the eigenvalues of a
matrix with respect to its entries, there exists an open set $U\subset
V$ containing $0$ such that $\forall\lambda\in U$, the solution
$\tilde{\underline{\Delta}}$ of the BP recursion (\ref{BPrec}) is
real, so that the corresponding spectral density in $U$ is equal to
$0$. This proves that the bulk of the spectrum of $H$ reaches $0$ at
$r=r_c=\sqrt{\rho({\rm B})}$, further justifying our choice for
the regularizer.

\section{Numerical results}
\label{sec:numerics}

\subsection{Synthetic networks}
\label{sec:SBM-test}

\begin{figure}[t]
\begin{center}
\includegraphics[width=0.99\linewidth]{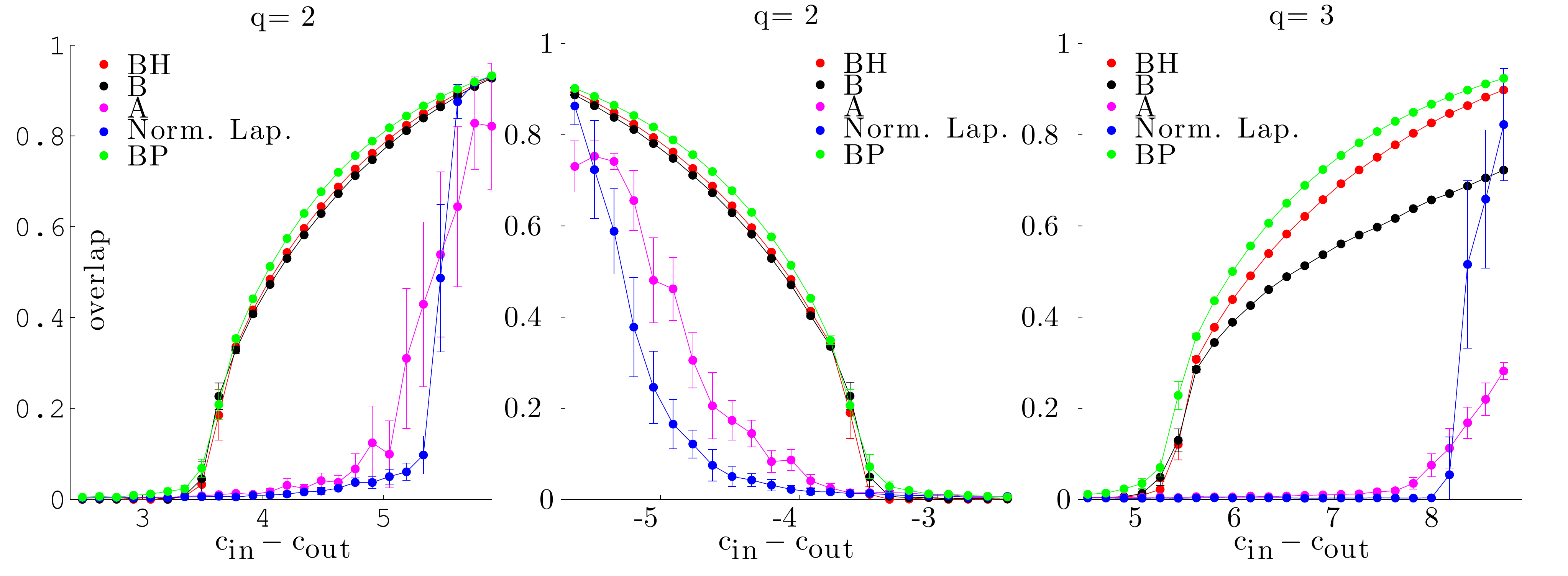}
\end{center}
\caption{Performance of spectral clustering applied to graphs of size
  $n=10^5$ generated from the the stochastic block model. Each point
  is averaged over $20$ such graphs. Left: assortative case with $q=2$
  clusters (theoretical transition at 3.46); middle: disassortative
  case with $q=2$ (theoretical transition at -3.46); right:
  assortative case with $q=3$ clusters (theoretical transition at
  5.20). For $q=2$, we clustered according to the signs of the
  components of the eigenvector corresponding to the second most
  negative eigenvalue of the Bethe operator. For $q=3$, we used
  k-means on the $3$ "negative" eigenvectors.  While both the standard
  adjacency (A) and symmetrically normalized Laplacian
  $(D^{-1/2}(D-A)D^{-1/2})$ approaches fail to identify clusters in a
  large relevant region, both the non-backtracking and the Bethe
  Hessian approaches identify clusters almost as well as using the
  more complicated belief propagation (BP) with oracle
  parameters. Note, however, that the Bethe Hessian systematically
  outperform the non-backtracking operator, at a smaller computational
  cost. \label{fig:SBM}}
\end{figure}
We now illustrate the efficiency of the algorithm for graphs generated
by the stochastic block model. Fig.~\ref{fig:SBM} shows the performance of standard spectral
clustering methods, as well as that of the belief propagation ($\rm BP$)
algorithm of \cite{decelle2011asymptotic}, believed to be
asymptotically optimal because it approximates the Bayes-optimal
estimator in a (conjectured) exact manner in large tree-like
graph. The performance is measured in terms of the overlap with the
true labeling, defined as \eq{
\label{overlap}
\left( \frac{1}{N} \sum_u \delta_{g_u , \tilde{g}_u} -
  \frac{1}{q}\right) \left/ \left( 1 - \frac{1}{q} \right) \right. \,
,  } 
where $g_u$ is the true group label of node $u$, and $\tilde{g}_u$ is
the label given by the algorithm, and we maximize over all
$q!$ possible permutation of the groups.
The Bethe Hessian systematically outperforms $\rm B$ and does almost
as well as $\rm BP$, which is a more complicated non-linear machine
learning algorithm, that we have run here assuming the knowledge of
"oracle parameters": the number of communities, their sizes, and the
matrix $p_{ab}$ \cite{decelle2011inference,decelle2011asymptotic}.  The Bethe Hessian, on the other hand can infer the
number of communities in the graph by counting the number of
negative eigenvalues, and does not need to be fed with the
interaction matrix.

\subsection{Real networks}
\label{sec:REAL}
We finally turn towards actual real graphs to illustrate
the performances of our approach in practical applications, and to show
that even if real networks are not generated by the stochastic block
model, the Bethe Hessian operator remains a useful tool. In Table \ref{sample-table} we give the overlap and the
number of groups to be identified for several networks commonly used as
benchmarks for community detection.  For each of these networks we
observed a large correlation to the ground truth, and at least equal
(and sometimes better) performances with respect to the non
backtracking operator. In all cases, the eigenvalues we have
considered lay in the negative part of the spectrum and are thus
clearly identifiable.

In particular, we find by counting the negative eigenvalues an
estimate of the number of clusters (just as
\cite{krzakala2013spectral} did for the non-backtracking matrix). It
is also interesting to note that our approach works not only in the
assortative case but also in the disassortative ones, for instance for
the word adjacency networks. A Matlab implementation to reproduce the
results of the Bethe Hessian for both real and synthetic network is
provided on the following
webpage: \url{http://mode_net.krzakala.org/}.

\begin{table}[t]
  \caption{Overlap for some commonly used benchmarks for community detection, computed using the signs of the second eigenvector for the
    networks with two communities, and using k-means for those with
    three and more communities, compared to the man-made group
    assignment. The non-backtracking operator
    detects communities in all these networks, with an overlap
    comparable to the performance of other spectral methods. The Bethe
    Hessian systematically either equals or outperforms the results
    obtained by the non-backtracking operator.
  }
  
\label{sample-table}
\begin{center}
\begin{tabular}{lcc}
\multicolumn{1}{c}{\bf PART}  &\multicolumn{1}{c}{\bf
  Non-backtracking \cite{krzakala2013spectral} } &\multicolumn{1}{c}{\bf Bethe Hessian}
\\ \hline \\
Polbooks ($q=3$)  \cite{adamic2005political}       &$0.742857$&$0.757143$  \\
Polblogs  ($q=2$)    \cite{lusseau2003bottlenose}           &$0.864157$& $0.865794$\\
Karate    ($q=2$)    \cite{zachary1977information}        &$1$ & $1$ \\
Football  ($q=12$) \cite{girvan2002community}       &$0.924111$ & $0.924111$\\
Dolphins    ($q=2$)   \cite{newman2006finding}          &$0.741935$& $0.806452$ \\
Adjnoun ($q=2$) \cite{Polbooks}    &$0.625000$ &$0.660714$ \\
\end{tabular}
\end{center}
\end{table}

\section{Conclusion and perspectives}
We have presented here a new approach to spectral clustering using the
Bethe Hessian and gave evidence that this approach combines the
advantages of standard sparse symmetric real matrices, with the
performances of the more involved non-backtracking operator, or the
use of the belief propagation algorithm with oracle parameters. This
answers the quest for a tractable non-parametric approach that
performs optimally in the stochastic bloc model. We invite the reader
to the demo file in matlab avaliable
at: \url{http://mode_net.krzakala.org/}.

Given the large impact and the wide use of spectral clustering methods
in many fields of modern science, we thus expect that our method will
have a significant impact on data analysis. In particular, our
approach can be straightforwardly generalized to other types of
spectral clustering problems, as for instance, those with real-valued
similarities $w_{ij}$ between vertices $i$ and $j$, without any price
in scalability (as opposed, for instance, to the non-backtracking
operator).

Another promising direction of investigation arises from the
observation that the cost function used in the graphical model of
section \ref{section:MODEL} can be generalized to any objective
function to be maximized, e.g. the modularity, or else. The Bethe
Hessian's negative eigenvalues at some carefully chosen regularization
parameter could therefore provide an approximate solution to this
maximization problem, giving a general spectral relaxation to some
known NP-hard problems.

\acknowledgments We would like to thank Federico Ricci-Tersenghi for
interesting discussions. This work has been supported in part by the
ERC under the European Union's 7th Framework Programme Grant Agreement
307087-SPARCS, by the Grant DySpaN of "Triangle de la Physique".

\bibliographystyle{plain}
\bibliography{../paper/mybib}

\end{document}